\input{aipcheck}

\documentclass[
    ,final            
  ]
  {aipproc}

\layoutstyle{6x9}

\usepackage{amsmath,amssymb}
\usepackage{bm}

\newcommand{\bra}[1]{\langle \, #1 \, |}
\newcommand{\ket}[1]{| \, #1 \, \rangle}

\usepackage[normalem]{ulem}  
\usepackage[dvips]{color} 

\renewcommand\sout{\bgroup \color{red} \ULdepth=-.5ex \ULset}

\begin{document}

\title{Compositeness of bound states \\
in chiral unitary approach}

\classification{14.20.--c, 11.30.Rd, 12.39.Fe}
\keywords      {Compositeness, chiral unitary approach, dynamically generated bound states}

\author{Tetsuo Hyodo}{
  address={Department of Physics, Tokyo Institute of Technology, 
Meguro 152-8551, Japan}
}

\author{Daisuke Jido}{
  address={Yukawa Institute for Theoretical Physics, 
Kyoto University, Kyoto 606--8502, Japan}
}

\author{Atsushi Hosaka}{
  address={Research Center for Nuclear Physics (RCNP),
Ibaraki, Osaka 567-0047, Japan}
}

\begin{abstract}
 We study the structure of dynamically generated bound states in the chiral unitary approach. The compositeness of a bound state is defined through the wavefunction renormalization constant in the nonrelativistic field theory.  We apply this argument to the chiral unitary approach and derive the relation between compositeness of the bound state and the subtraction constant of the loop integral. The compositeness condition is fairly compatible with the natural renormalization scheme, previously introduced in a different context.
\end{abstract}

\maketitle


\section{Introduction}

One of the recent topics in hadron spectroscopy is to reveal the internal structure of hadrons. In particular, hadron resonances located near two-hadron threshold are considered to be dominated by the hadronic molecule structure which is the two-hadron quasibound state driven by an inter-hadron force. For instance, the baryon resonance $\Lambda(1405)$ and the scalar mesons $a_0(980)$ and $f_0(980)$ can be dynamically generated by the low energy $s$-wave chiral interaction, in the framework of the chiral unitary approach~\cite{Kaiser:1995eg,Oset:1998it,Oller:2000fj,Lutz:2001yb}, which also well describes many hadron resonances in a reasonable way~\cite{Hyodo:2006yk,Hyodo:2006kg}. 

It was however pointed out that certain choice of the cutoff parameter may introduce the seed of the resonance, even with the leading order chiral interaction which does not contain the pole term contributions~\cite{Hyodo:2008xr}. To exclude such contribution from the loop function, the natural renormalization scheme was proposed, which enables us to pin down the origin of the generated resonance qualitatively. As a next step, it is desired to define a quantitative measure of compositeness in the chiral unitary approach. In this report we attempt to calculate the compositeness of the bound state, defined through the field renormalization constant $Z$~\cite{Weinberg:1962hj,Weinberg:1965zz}, in the chiral unitary approach.

\section{Compositeness of bound states}

We first consider the two-body scattering system with one bound state in a nonrelativistic field theory, and define the field renormalization constant following Ref.~\cite{Weinberg:1965zz}. We decompose the full Hamiltonian $\mathcal{H}$ into the free part $\mathcal{H}_0$ and the interaction $V$. We assume that the Hilbert space of $\mathcal{H}_0$ consists of the bare bound state $\ket{B_0}$ and the two-body scattering states labeled by their momenta as
\begin{align}
    1 
    &= \ket{B_0}\bra{B_0} + \int d\bm{k} \ket{\bm{k}}\bra{\bm{k}}
    .
    \label{eq:complete}
\end{align}
In the same way, the eigenstates of the full hamiltonian $\mathcal{H}$ are given by the physical bound state $\ket{B}$ with the binding energy $B>0$ and the continuum states $\ket{\bm{k},\text{full}}$ as
\begin{align}
    1 
    &= \ket{B}\bra{B} + \int d\bm{k} \ket{\bm{k},\text{full}}\bra{\bm{k},\text{full}}
    .
    \label{eq:completefull}
\end{align}
Now we define the field renormalization constant $Z$ as the overlap of the bare state $\ket{B_0}$ and the physical bound state as
\begin{align}
    Z
    & \equiv |\bra{B_0}B\, \rangle |^2 ,
    \label{eq:Zdef}
\end{align}
which takes the value $0\leq Z\leq 1$. Since $\ket{B_0}$ is considered to be an elementary particle, $Z$ expresses the degree of ``elementarity" of the physical bound state. The quantity $1-Z$ can then be interpreted as the ``compositeness" of the bound state;
\begin{align}
    1-Z
    &= \int d\bm{k}\frac{|\bra{\bm{k}}V\ket{B}|^2}{[E(\bm{k})+B]^2} 
    = 4\pi\sqrt{2\mu^3}
    \int_{0}^{\infty} dE\frac{\sqrt{E}
    |G_W(E)|^2}{(E+B)^2} ,
    \label{eq:1mZexact}
\end{align}
where $\mu$ is the reduced mass and we consider the $s$-wave bound state whose vertex form factor $\bra{\bm{k}}V\ket{B}\equiv G_W(E)$ has no angular dependence. On the other hand, Low's scattering equation can be obtained by using the formal solution of the Lippmann-Schwinger equation and the complete set of the full Hamiltonian~\eqref{eq:completefull} as~\cite{Weinberg:1965zz}
\begin{align}
    t(E)
    =&v
    +
    \frac{|G_W(E)|^2}{E+B} 
    +4\pi\sqrt{2\mu^3}\int_0^{\infty} dE^{\prime}
    \frac{\sqrt{E^{\prime}}|t(E^{\prime})|^2}{E-E^{\prime}+i\epsilon} ,
    \label{eq:Wamp1}
\end{align}
where $v$ is the matrix element of the interaction Hamiltonian $V$ by the scattering states. The second term of this equation is a part of the integrand of Eq.~\eqref{eq:1mZexact}. Therefore, the compositeness can be expressed by the scattering amplitude
\begin{align}
    1-Z
    =&4\pi\sqrt{2\mu^3}
    \int_{0}^{\infty} dE \frac{\sqrt{E}}{E+B}
    \Biggl[
    t(E)-v 
    -
    4\pi\sqrt{2\mu^3}\int_{0}^{\infty} dE^{\prime}
    \frac{\sqrt{E^{\prime}}|t(E^{\prime})|^2}{E-E^{\prime}+i\epsilon}
    \Biggr] .
    \label{eq:compositenessNRexact}
\end{align}
Although the scattering amplitude $t(E)$ is in general complex, the imaginary part vanishes due to the optical theorem. 

Let us consider the limit $B\to 0$. For small $B$, the integrand in Eq.~\eqref{eq:compositenessNRexact} is dominated by the bound state pole term and the vertex form factor can be approximated by the coupling constant $g_W \equiv G_W(E=-B)$~\cite{Weinberg:1965zz}. Thus, in the small binding limit, we obtain
\begin{align}
    1-Z
    \approx
    &
    2\pi^2  
    \sqrt{2\mu^3} \frac{g_W^2}{\sqrt{B}} .
    \label{eq:compositenessNR}
\end{align}
This is the famous result of Ref.~\cite{Weinberg:1965zz}, which connects the compositeness of the bound state with the coupling constant $g_W$ and the binding energy $\sqrt{B}$. 

Note that Eq.~\eqref{eq:compositenessNR} is an approximated form valid only for the small $B$, but the result is model independent because the explicit form of the interaction $V$ is not relevant. All the information of the interaction is embedded in the binding energy and coupling constant. In contrast, Eq.~\eqref{eq:compositenessNRexact} is the exact expression of $1-Z$ which can be applied to the bound state with an arbitrary binding energy. The expression however include the matrix element of the interaction $V$, so the result depends on the particular choice of the interaction Hamiltonian. In the following, we utilize the model-independent form of Eq.~\eqref{eq:compositenessNR} for the analysis of the bound state in the chiral unitary approach.\footnote{The completeness of the full hamiltonian is not always guaranteed with the energy-dependent interaction $V$, as in the case of chiral unitary approach. The use of the Low's equation is valid when the eigenstates of the full Hamiltonian span the complete and orthonormal basis, while the result of Eq.~\eqref{eq:compositenessNR} can be derived without using the completeness of the full Hamiltonian~\cite{Weinberg:1965zz}.}

\section{Compositeness in chiral unitary approach}

Here we describe the same $s$-wave scattering system in the chiral unitary approach. The low energy theorem gives the Weinberg-Tomozawa (WT) interaction
\begin{align}
    V(W)
    =& C(W-M) 
     \label{eq:WTinteraction} ,
\end{align}
where $M$ is the mass of the target hadron, $W$ is the total center of mass energy, and $C$ is a real-valued constant with mass dimension minus two. The scattering amplitude can be obtained by the N/D method~\cite{Oller:2000fj} as\begin{align}
    T(W)
    &=
    \frac{1}{1-V(W)G(W;a)}V(W)
    \label{eq:amplitude} .
\end{align}
The loop function $G(W;a)$ with the dimensional regularization as
\begin{align}
    G(W;a)=&\frac{2M}{(4\pi)^{2}}
    \Biggl(a
    +\frac{m^{2}-M^{2}+W^2}{2W^2}\ln\frac{m^{2}}{M^{2}}
    +\frac{\bar{q}(W)}{W}
    \Bigl\{\ln[W^2-(M^{2}-m^{2})+2W\bar{q}(W)] \nonumber\\
    &+\ln[W^2+(M^{2}-m^{2})+2W\bar{q}(W)] 
    -\ln[-W^2+(M^{2}-m^{2})+2W\bar{q}(W)]
    \nonumber\\
    &
    -\ln[-W^2-(M^{2}-m^{2})+2W\bar{q}(W)]
    \Bigr\}\Biggr) ,
    \label{eq:Gfn}
\end{align}
where $\bar{q}(W)=\sqrt{[W^2-(M-m)^2][W^2-(M+m)^2]}/2W$. The ultraviolet divergence is removed and the finite part is specified by the subtraction constant $a$.

Let us calculate the compositeness in Eq.~\eqref{eq:compositenessNR} in this framework. Apart from the masses of the particles, the amplitude~\eqref{eq:amplitude} contains two parameters, the coupling strength $C$ and the cutoff parameter $a$. The condition for having a bound state at $M_B$ reads
\begin{align}
    1-C(M_B-M)G(M_B;a)
    =0 .
    \label{eq:WTcond}
\end{align}
which relates $C$, $M_B$ and $a$, so we can characterize the system by ($M_B$, $a$), instead of ($C$, $a$). The coupling strength of the bound state to the scattering state $g^2$ can be calculated from the residue of the pole as
\begin{align}
    [g(M_B;a)]^2 
    =& \lim_{W\to M_B}(W-M_B)T(W)
    =-
    \frac{M_B-M}
    {
    G(M_B;a)
    +
    (M_B-M)
    G^{\prime}(M_B)} .
    \label{eq:WTcoupling}
\end{align}
It is important to note that the coupling constant depends on the subtraction constant. If the interaction \eqref{eq:WTinteraction} is energy independent, this coupling constant is simply given by the derivative of the loop function, which is independent of the subtraction constant. Taking into account the normalization difference of two approaches, we obtain the compositeness of the bound state in the chiral unitary approach as
\begin{align}
    1-Z
    &= 
    \frac{M|\bar{q}(M_B)|}{8\pi M_B(M+m-M_B)}
    [g(M_B;a)]^2 
    \label{eq:compositenessChU}
\end{align}
For the given mass of the bound state $M_B$ and the subtraction constant $a$, we can calculate the compositeness using Eqs.~\eqref{eq:WTcoupling} and \eqref{eq:compositenessChU}.

\section{Numerical results}

In Ref.~\cite{Hyodo:2008xr}, the natural subtraction constant $a_{\text{natural}}$ was introduced to exclude CDD pole contributions from the loop function, based on the consistency with the property of the sensible loop function and the matching with the chiral low energy interaction. If this is the case, we expect that the generated bound state should be a composite particle, leading to $Z\sim 0$. Here we study this numerically.

\begin{figure}[tbp]
\includegraphics[width=7cm,clip]{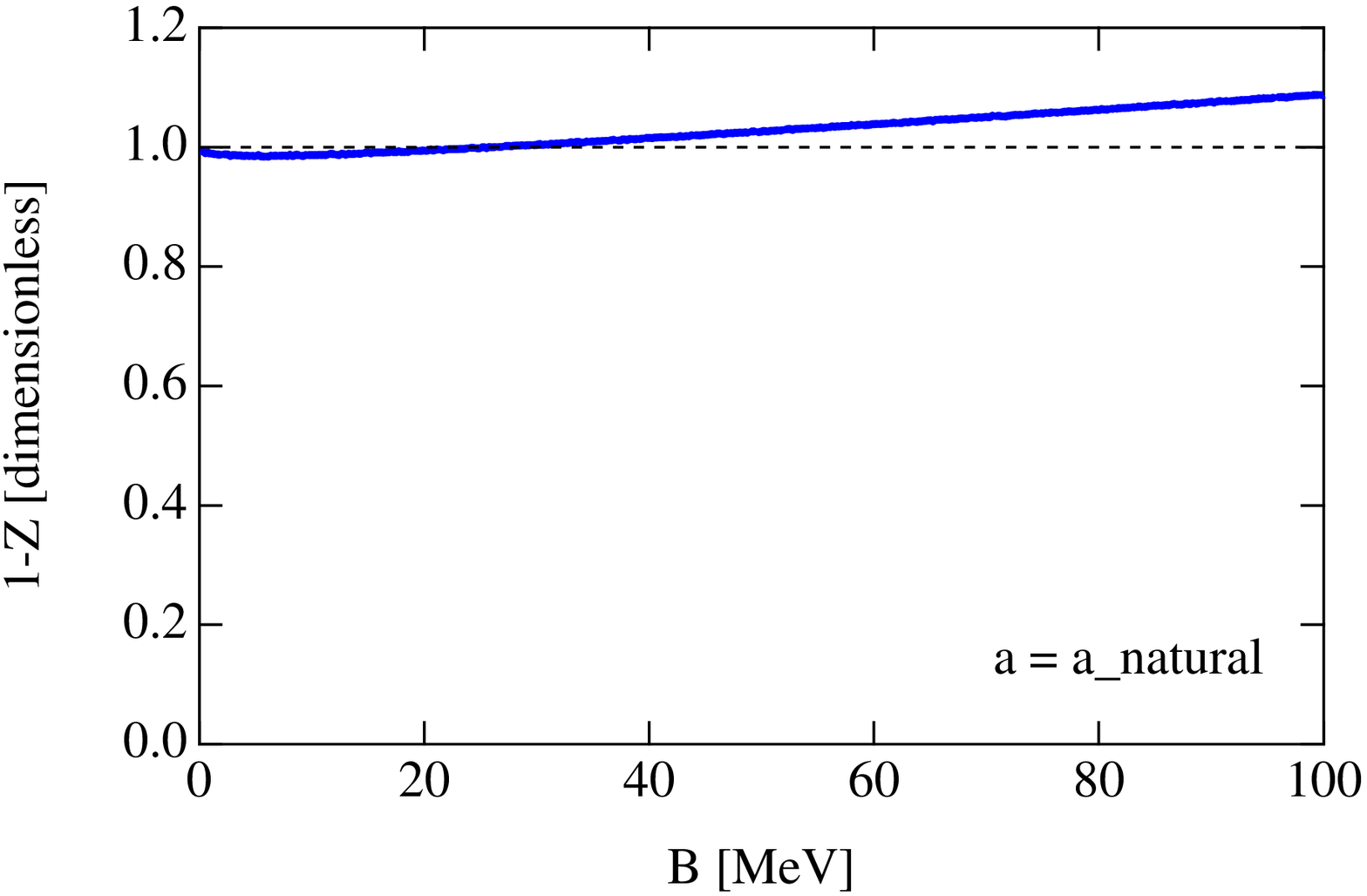}
\includegraphics[width=7cm,clip]{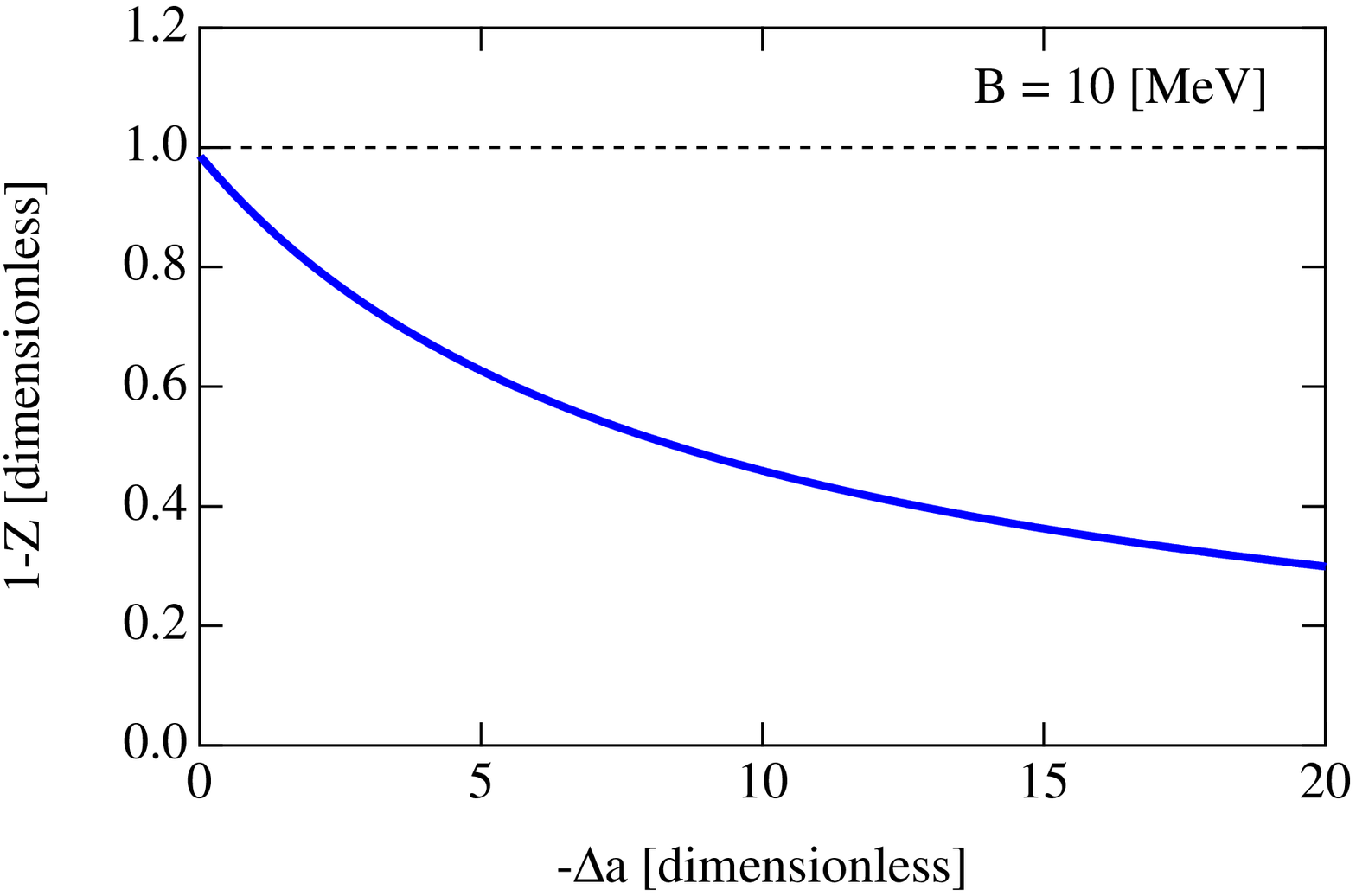}
\caption{\label{fig:1mZ} The compositeness $1-Z$ as a function of the binding energy $B=M+m-M_B$ with the natural subtraction constant (Left) and of the deviation from the natural value $-\Delta a\equiv a_{\text{natural}}-a$ with the binding energy $B=10$ MeV (Right). Particle masses are set to be $m=495$ MeV and $M=939$ MeV.}
\end{figure}

In the left panel of Fig.~\ref{fig:1mZ}, we plot the compositeness $1-Z$ as a function of the binding energy $B=M+m-M_B$ with the natural subtraction constant $a_{\text{natural}}$. As expected, we clearly see that $Z\sim 0$ for small binding energy. For $B\gtrsim 25$ MeV, the quantity $1-Z$ exceeds $1$, which may indicate the breakdown of the small binding approximation in Eq.~\eqref{eq:compositenessNR}.

Next we study the compositeness by varying the subtraction constant from the natural value, fixing the binding energy to be a constant. In this case, the coupling strength $C$ is adjusted according to Eq.~\eqref{eq:WTcond}. In the right panel of Fig.~\ref{fig:1mZ}, we show $1-Z$ as a function of $-\Delta a\equiv a_{\text{natural}}-a$ with the binding energy $B=10$ MeV. We find that the compositeness decreases when the deviation from $a_{\text{natural}}$ becomes large. This is the effect of the seed of the resonance, introduced through the cutoff parameter in the loop function. It is also instructive to consider the pole mass in the effective interaction $M_{\text{eff}}=M+(4\pi)^2/(2M C \Delta a)$. It can be shown that $Z$ approaches to unity in the limit $M_{\text{eff}}\to M_B$. This means that the bound state is dominated by the elementary component, if the effective mass appears close to the physical bound state. 

\section{Summary}

The compositeness of the bound state is studied in the chiral unitary approach.  We define the compositeness using the field renormalization constant, and express it in terms of the scattering amplitude. We note that the compositeness depends on the explicit form of the interaction Hamiltonian $V$. We analyze the bound state in the chiral unitary approach, and derive the analytic form of the compositeness in the small binding limit. The energy dependence of the chiral interaction causes the subtraction constant dependence of the compositeness. Varying the value of the subtraction constant, we find that the bound state in the natural renormalization scheme shows $Z\sim 0$, while the large deviation from the natural value indicates the substantial component from the elementary particle. These observations support the analysis of the origin of the resonances using the natural renormalization condition~\cite{Hyodo:2008xr}. The formulation of the compositeness in the relativistic field theory is of relevance and will be summarized elsewhere~\cite{Compositeness}.






\begin{theacknowledgments}
  T.H. thanks the support from the Global Center of Excellence Program by MEXT, Japan through the Nanoscience and Quantum Physics Project of the Tokyo Institute of Technology. This work was partly supported by the Grant-in-Aid for Scientific Research from MEXT and JSPS (Nos. 22740161, 22105507 and 21840026).
\end{theacknowledgments}	





\IfFileExists{\jobname.bbl}{}
 {\typeout{}
  \typeout{******************************************}
  \typeout{** Please run "bibtex \jobname" to optain}
  \typeout{** the bibliography and then re-run LaTeX}
  \typeout{** twice to fix the references!}
  \typeout{******************************************}
  \typeout{}
 }

\end{document}